\documentclass[twocolumn, tighten]{AASTeX62}

\usepackage{url}
\usepackage{amsmath}
\usepackage{multirow}
\usepackage{booktabs}
\usepackage{hyperref}
\usepackage{latexsym}

\received{August 3, 2019}
\revised{August 20, 2019}
\accepted{August 21, 2019}

\shorttitle{1987A-like Type II SN~2018hna}
\shortauthors{Singh et al. 2019}

\begin{document}

\title{SN~2018hna: 1987A-like supernova with a signature of shock breakout}

\correspondingauthor{Avinash Singh}
\email{avinash21292@gmail.com, avinash.singh@iiap.res.in}

\author[0000-0003-2091-622X]{Avinash Singh}
\affiliation{Indian Institute of Astrophysics, Koramangala 2nd Block, Bengaluru 560034, India}
\affiliation{Joint Astronomy Programme, Department of Physics, Indian Institute of Science, Bengaluru 560012, India}

\author{D.K. Sahu}
\affiliation{Indian Institute of Astrophysics, Koramangala 2nd Block, Bengaluru 560034, India}

\author[0000-0003-3533-7183]{G.C. Anupama}
\affiliation{Indian Institute of Astrophysics, Koramangala 2nd Block, Bengaluru 560034, India}

\author[0000-0001-7225-2475]{Brajesh Kumar}
\affiliation{Indian Institute of Astrophysics, Koramangala 2nd Block, Bengaluru 560034, India}

\author{Harsh Kumar}
\altaffiliation{LSSTC DSFP Fellow}
\affiliation{Physics Department, Indian Institute of Technology Bombay, Powai, Mumbai 400076, India}

\author[0000-0001-9456-3709]{Masayuki Yamanaka}
\affiliation{Okayama Observatory, Kyoto University, 3037-5 Honjo, Kamogata-cho, Asakuchi, Okayama 719-0232, Japan}
\affiliation{Hiroshima Astrophysical Science Center, Hiroshima University, Higashi-Hiroshima, Hiroshima 739-8526, Japan}

\author{Petr V. Baklanov}
\affiliation{National Research Center "Kurchatov Institute" - ITEP, ul. Bolshaya Cheremushkinskaya 25, Moscow 117218, Russia}
\affiliation{National Research Nuclear University MEPhI, Kashirskoe sh. 31, Moscow 115409, Russia}

\author[0000-0001-8537-3153]{Nozomu Tominaga}
\affiliation{Department of Physics, Faculty of Science and Engineering, Konan University, 8-9-1 Okamoto, Kobe, Hyogo 658-8501, Japan}
\affiliation{Kavli Institute for the Physics and Mathematics of the Universe (WPI), The University of Tokyo Institutes for Advanced Study, The University of Tokyo, 5-1-5 Kashiwanoha, Kashiwa, Chiba 277-8583, Japan}

\author{Sergei I. Blinnikov}
\affiliation{National Research Center "Kurchatov institute", Institute for Theoretical and Experimental Physics (ITEP), 117218 Moscow, Russia}
\affiliation{Kavli Institute for the Physics and Mathematics of the Universe (WPI), The University of Tokyo Institutes for Advanced Study, The University of Tokyo, 5-1-5 Kashiwanoha, Kashiwa, Chiba 277-8583, Japan}
\affiliation{Sternberg Astronomical Institute, M.V. Lomonosov Moscow State University, Universitetski pr. 13, 119234 Moscow, Russia}

\author[0000-0003-2611-7269]{Keiichi Maeda}
\affiliation{Department of Astronomy, Kyoto University, Kitashirakawa-Oiwake-cho, Sakyo-ku, Kyoto 606-8502, Japan}

\author{Anirban Dutta}
\affiliation{Indian Institute of Astrophysics, Koramangala 2nd Block, Bengaluru 560034, India}

\author[0000-0002-6112-7609]{Varun Bhalerao}
\affiliation{Physics Department, Indian Institute of Technology Bombay, Powai, Mumbai 400076, India}

\author{Ramya M. Anche}
\affiliation{Indian Institute of Astrophysics, Koramangala 2nd Block, Bengaluru 560034, India}

\author[0000-0002-3927-5402]{Sudhanshu Barway}
\affiliation{Indian Institute of Astrophysics, Koramangala 2nd Block, Bengaluru 560034, India}

\author{Hiroshi Akitaya}
\affiliation{Hiroshima Astrophysical Science Center, Hiroshima University, Higashi-Hiroshima, Hiroshima 739-8526, Japan}

\author{Tatsuya Nakaoka}
\affiliation{Department of Physical Science, Hiroshima University, Kagamiyama 1-3-1, Higashi-Hiroshima 739-8526, Japan}

\author{Miho Kawabata}
\affiliation{Department of Physical Science, Hiroshima University, Kagamiyama 1-3-1, Higashi-Hiroshima 739-8526, Japan}
\affiliation{Department of Astronomy, Kyoto University, Kitashirakawa-Oiwake-cho, Sakyo-ku, Kyoto 606-8502, Japan}

\author{Koji S Kawabata}
\affiliation{Hiroshima Astrophysical Science Center, Hiroshima University, Higashi-Hiroshima, Hiroshima 739-8526, Japan}

\author{Mahito Sasada}
\affiliation{Hiroshima Astrophysical Science Center, Hiroshima University, Higashi-Hiroshima, Hiroshima 739-8526, Japan}

\author{Kengo Takagi}
\affiliation{Department of Physical Science, Hiroshima University, Kagamiyama 1-3-1, Higashi-Hiroshima 739-8526, Japan}

\author{Hiroyuki Maehara}
\affiliation{Okayama Branch Office, Subaru Telescope, National Astronomical Observatory of Japan, NINS, Kamogata, Asakuchi, Okayama 719-0232, Japan}

\author{Keisuke Isogai}
\affiliation{Okayama Observatory, Kyoto University, 3037-5 Honjo, Kamogata-cho, Asakuchi, Okayama 719-0232, Japan}

\author{Masaru Kino}
\affiliation{Okayama Observatory, Kyoto University, 3037-5 Honjo, Kamogata-cho, Asakuchi, Okayama 719-0232, Japan}

\author{Kenta Taguchi}
\affiliation{Department of Astronomy, Kyoto University, Kitashirakawa-Oiwake-cho, Sakyo-ku, Kyoto 606-8502, Japan}

\author[0000-0002-3933-7861]{Takashi Nagao}
\affiliation{European Southern Observatory, Karl-Schwarzschild-Str. 2, 85748 Garching b. M\"{u}nchen, Germany}

\begin{abstract}
High cadence ultraviolet, optical and near-infrared photometric and low-resolution spectroscopic observations of the peculiar Type II supernova (SN) 2018hna are presented. The early phase multiband light curves exhibit the adiabatic cooling envelope emission following the shock breakout up to $\sim$\,14 days from the explosion. SN~2018hna has a rise time of $\sim$\,88 days in the $V$-band, similar to SN~1987A. A $\rm^{56}Ni$ mass of $\sim$\,0.087\,$\pm$\,0.004 $\rm M_{\odot}$ is inferred for SN~2018hna from its bolometric light curve. Hydrodynamical modelling of the cooling phase suggests a progenitor with a radius $\sim$\,50 $\rm R_{\odot}$, a mass of $\sim$\,14--20 $\rm M_{\odot}$ and an explosion energy of $\sim$\,1.7--2.9$\rm \times$\,$\rm 10^{51}\ erg$. The smaller inferred radius of the progenitor than a standard red supergiant is indicative of a blue supergiant progenitor of SN~2018hna. A sub-solar metallicity ($\sim$\,0.3 $\rm Z_{\odot}$) is inferred for the host galaxy UGC~07534, concurrent with the low-metallicity environments of 1987A-like events.
\end{abstract}


%-----------------------------------------------------------------------------%
\section{Introduction}\label{sec:intro}
%-----------------------------------------------------------------------------%

Core-collapse supernovae (CCSNe) result from the gravitational collapse of stars with a Zero-Age Main Sequence (ZAMS) mass $\gtrsim$\,8 $\rm M_{\odot}$ \citep{2003heger}. Type II SNe result from stars that retain their hydrogen envelope at the time of explosion. Theoretical \citep{1971grassberg} and observational \citep{2009smartt} studies of Type II SNe indicate a Red Super-Giant (RSG) progenitor. However, the nearest naked-eye supernova in the last four centuries, SN~1987A showed that a Blue Super-Giant (BSG) can also be a progenitor of Type II SNe. The presence of a slow rise to the maximum ($\sim$\,85\,--\,100 d) is the distinguishing feature in the light curves of Type II SNe resulting from a BSG \citep{2011kleiser}. However, extended RSGs can also result in a slow rising Type II SN if they synthesize a substantial amount of $\rm ^{56}Ni$ ($\sim$\,0.2 $\rm M_{\odot}$) in the explosion (e.g.\ SN~2004ek, \citealp{2016taddia}). Theoretical models have shown that fast-rotation, low-metallicity and/or interaction in a binary system can indeed lead to the explosion of BSGs as SNe \citep{1992podsiadlowski}.

In the vast expanse of data on Type II SNe, only a handful of SNe have shown similarities to SN~1987A \citep{1990hamuy,1995pun}. These are SN~1998A \citep{2005pastorello}, SN~2000cb and SN~2005ci \citep{2011kleiser}, SN~2006V and SN~2006au \citep{2012taddia}, SN~2009E \citep{2012pastorello}, SN~2009mw \citep{2016takats} and SN~Refsdal \citep{2016rodney,2016kelly}. These will be referred to as 1987A-like events, whereas, Type II events with an RSG progenitor will be referred to as normal Type II SNe. Analysis of 1987A-like events have shown that these arise from rather compact progenitors (BSG, R\,$<$\,100 $\rm R_{\odot}$) with a higher ZAMS mass, higher explosion energies ($>$\,$\rm 10^{51}$ erg) and higher synthesized $\rm ^{56}Ni$ ($\sim$\,0.1 $\rm M_{\odot}$) when compared to normal Type II SNe \citep{2012pastorello,2016taddia}.

The collapse of the core in CCSNe is succeeded by a shock wave resulting from the rebound of the in-falling matter on the neutron-degenerate proto-neutron star (collapsed core). A fraction of the energy lost during this collapse is transferred via neutrino heating to the outer ejecta. The shock accelerates towards the surface of the star, releasing energy in X-ray and UV \citep{1977falk} due to its high temperature ($\rm 10^5-10^6\ K$, \citealp{1992ensman}). This short-lived ($\sim$1000s in RSGs) phase is labelled as the \lq \lq shock\rq \rq\ breakout and marks the first electromagnetic signature of a SN explosion.

The cooling emission from the heated envelope as a result of the shock breakout has been seen only in a few Type II SNe such as SN~1987A \citep{1992ensman}, SNLS-04D2dc \citep{2009tominaga}, SN~2010aq \citep{2010gezari} etc. The signature of a shock breakout is generally more prominent in Type IIb SNe with extended envelopes, e.g. SN~1993J \citep{1994richmond} and SN~2011fu \citep{2013brajesh}. The thermal emission from the heated ejecta peaks in the UV spanning a few hours to a couple of days \citep{2010nakar}. The shock breakout may be delayed due to the presence of a circumstellar wind as the shock not only has to escape the outer envelope of the SN ejecta, but also the dense wind surrounding it \citep{2013ofek}. During the breakout, the energy released scales with the progenitor radius and hence the detection of an early emission helps in directly tracing the progenitor properties. 

SN~2018hna was discovered by Koichi Itagaki on 2018 October 23.9 UT (JD 2458414.3) in the galaxy UGC~07534. SN~2018hna lies 31\arcsec E and 30\arcsec N from the nucleus of the host galaxy. A spectrum obtained by Hiroshima One-shot Wide-field POLarimeter (HOWPol, \citealp{2008kawabata}) on 2018 October 24.8 UT, displayed a prominent P-Cygni profile of H\,$\rm \alpha$ categorizing it as a Type II SN (TNS\,\#\,2933).

The photometric and spectroscopic evolution of SN~2018hna is presented in this Letter, and discussed in the context of 1987A-like events.

\begin{figure*}
\centering
\resizebox{\hsize}{!}{\includegraphics{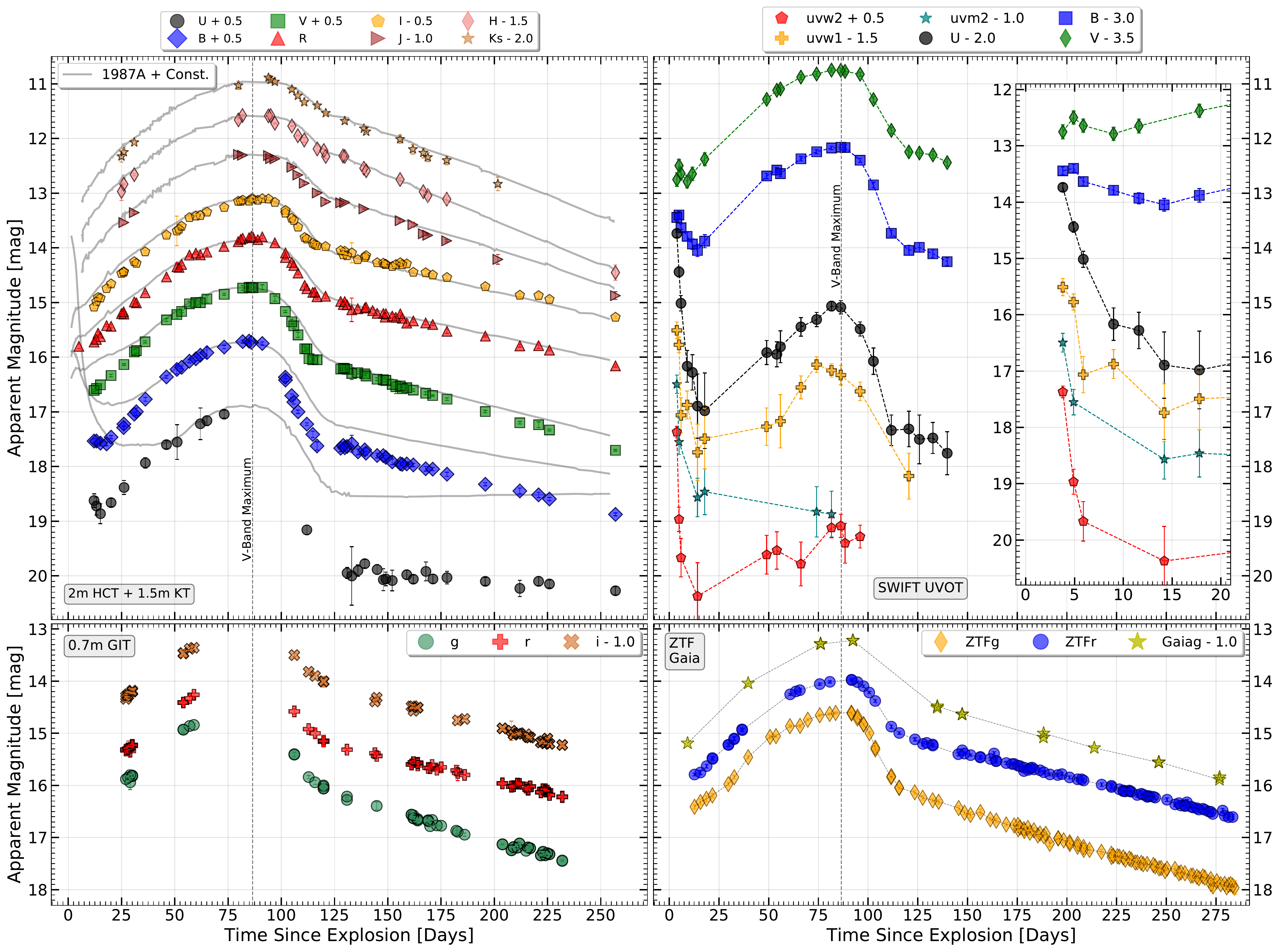}}
\caption{Apparent light curves (LCs) of SN~2018hna. The LCs of SN~1987A were shifted to match the maximum of SN~2018hna. Offsets have been applied for clarity.}
\label{fig:apparentlc}
\end{figure*}

%-----------------------------------------------------------------------------%
\section{Data Acquisition}\label{sec:data}
%-----------------------------------------------------------------------------%

Optical photometric ($UBVRI$) and spectroscopic observations of SN~2018hna using the 2-m Himalayan Chandra Telescope (HCT), Indian Astronomical Observatory (IAO), Hanle, India began on 2018 October 31.9 UT (JD 2458423.4). The recently installed robotic 0.7-m GROWTH\footnote{Global Relay of Observatories Watching Transients Happen (\url{http://growth.caltech.edu})} - India telescope (GIT) at IAO followed up SN~2018hna in the SDSS $g'r'i'$ filters starting 2018 November 15.9 UT. SN~2018hna was also monitored with the 1.5-m Kanata Telescope in the optical and the near-infrared using the HOWPol and the Hiroshima Optical and Near-InfraRed camera (HONIR; \citealp{2014akitaya}). Few spectra were also obtained using the Kyoto Okayama Optical Low-dispersion Spectrograph with Optical-Fiber Integral Field Unit (KOOLS-IFU, \citealp{2019matsubayashi}) mounted on the 3.8-m Seimei telescope, at the Okayama Observatory. Data reduction was performed using standard reduction procedures described in \citet{2018avinash}. The follow-up of SN~2018hna is supplemented by data from the public archives of Zwicky Transient Facility (ZTF; \citealp{2019bellm}) in $g$ and $r$ bands, from Lasair \citep{2019smith}, and Gaia \citep{2018gaia} in $g$-band.

The Neil Gehrels Swift Observatory \citep{2004gehrels} monitored SN~2018hna with the ultraviolet Optical Telescope (UVOT; \citealp{2005roming}) beginning 2018 October 23.6 UT. Data reduction was performed using the UVOT data analysis software in HEASOFT \citep[see][]{2018brajesh}.

%-----------------------------------------------------------------------------%

%-----------------------------------------------------------------------------%
\section{Photometric Evolution}\label{sec:photevol}
%-----------------------------------------------------------------------------%
The ultraviolet, optical and near-infrared light curves (LCs) of SN~2018hna are shown in Figure~\ref{fig:apparentlc}. The slow rise to the maximum and the broad peak resemble the peculiar Type II SN~1987A. Due to the immediate follow up with \textit{Swift}, the early LCs in the optical and the ultraviolet bands show a noticeable decline in brightness during the initial phase, followed by a rise to maximum. This drop can be explained as a result of the rapid cooling of the photosphere proceeding the shock breakout, and was also seen in SN~1987A. Hydrodynamic modelling of the early LC constrained the explosion date to be JD 2458411.3, 3 days prior to discovery (see Section~{\ref{sec:coolenv}}). The peak in the $V$-band LC of SN~2018hna at $\sim$\,87.5~d is adopted as the epoch of maximum. This is similar to SN~1987A ($\sim$\,86 days). 
A Galactic reddening of $E(B-V)$\,=\,0.009$\pm$0.001 mag along the direction of SN~2018hna was obtained from the dust-extinction map of \citet{2011schlafly}. This is consistent with the absence of interstellar \ion{Na}{1} D absorption in the spectra of SN~2018hna. No trace of \ion{Na}{1} D absorption is seen at the redshift of the host galaxy, concurrent with the high offset of SN~2018hna from the center of the host. A net reddening of $E(B-V)$\,=\,0.009$\pm$0.001 mag is adopted assuming no host extinction. A distance modulus of $\rm \mu$ = 30.52\,$\pm$\,0.29 mag (12.82\,$\pm$\,2.02 Mpc) is estimated for the host galaxy UGC~07534 from the mean of the Hubble flow distances provided in NASA Extragalactic Database (NED\footnote{\url{https://ned.ipac.caltech.edu}}) and the Tully-Fisher relation \citep{2013akarachentsev}.

%-----------------------------------------------------------------------------%
\section{Spectroscopic Evolution}\label{sec:specevol}
%-----------------------------------------------------------------------------%

\begin{figure*}
\centering
\resizebox{\hsize}{!}{\includegraphics{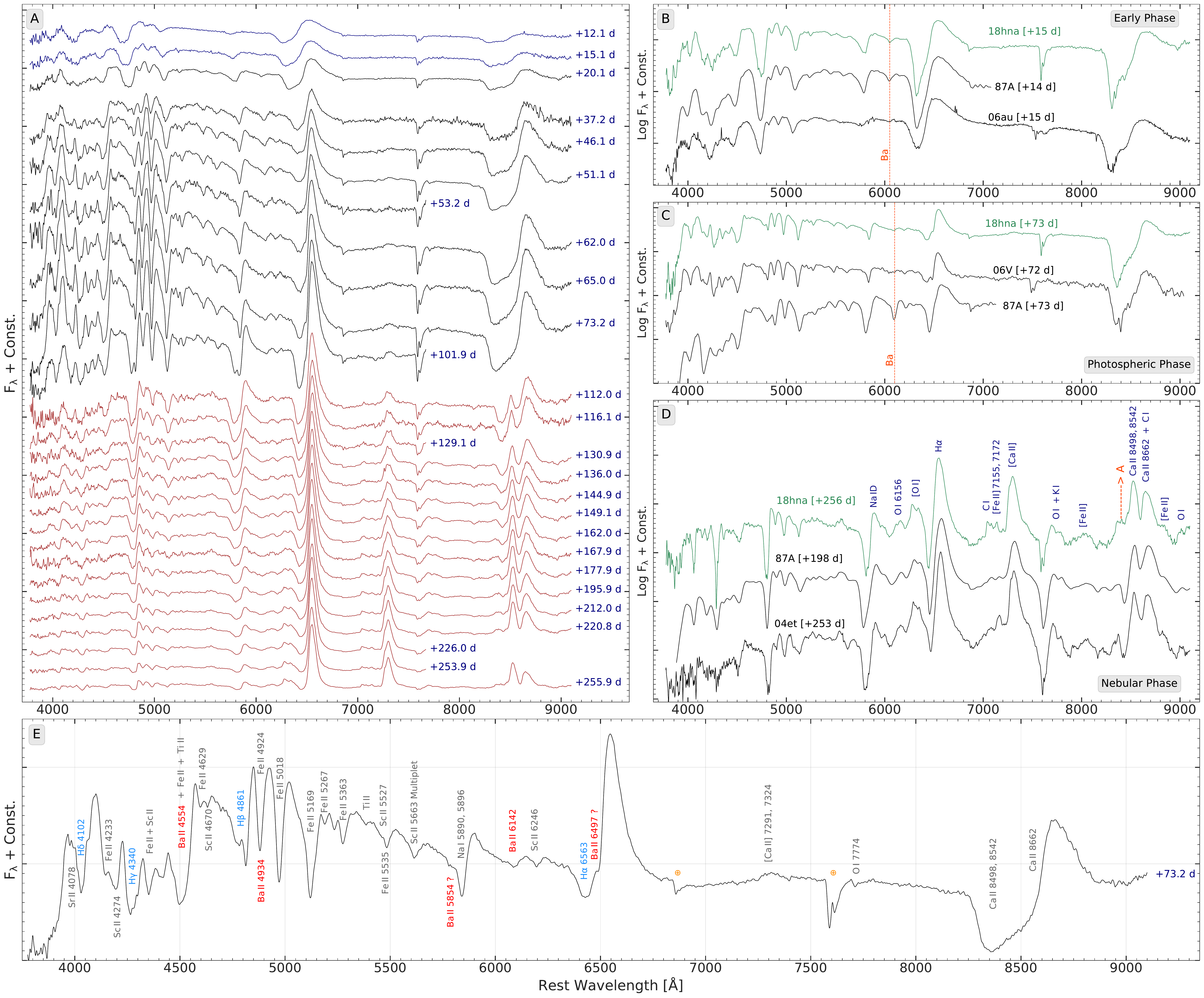}}
\caption{$Panel\ A$: Spectroscopic sequence of SN~2018hna. The three different colors depict the cooling envelope (blue), photospheric (black) and nebular phase (red) of the SN. $Panels\ B,\ C\ \&\ D$: Comparison of SN~2018hna with 1987A-like events. $Panel\ E$: Identification of lines in the spectrum of $\sim$\,73 d.}
\label{fig:specplat}
\end{figure*}

The spectral evolution of SN~2018hna is shown in Figure~\ref{fig:specplat}A. The first spectrum of SN~2018hna obtained at $\sim$\,12 d, shows strong, broad P-Cygni profiles of the Balmer lines, \ion{Fe}{2} features, \ion{Ca}{2} NIR triplet, and weak \ion{Ba}{2} 4554 \AA. \ion{He}{1} 5876 \AA\ was identified in the early ($\sim 3$ days from explosion) spectrum of SN~1987A. However, the feature around 5740 \AA\ is identified as \ion{Na}{1}\,D instead of \ion{He}{1} 5876 \AA\ due to the low color temperature ($<$\,8000 K) inferred from the Spectral Energy Distribution (SED). 

As the SN evolved towards the maximum, features of \ion{Na}{1} D, \ion{Fe}{2}, \ion{Ca}{2}, \ion{Ba}{2}, \ion{Sc}{2}, \ion{Ti}{2} and \ion{Sr}{2} became more prominent. The line identification in the spectrum of $\sim$\,73 d is shown in Figure~\ref{fig:specplat}E. The features of \ion{Ba}{2} (i.e. 4554 and 6142 \AA), which are characteristic of 1987A-like events \citep{1995mazzali} are also evident in the spectra ($>$\,15 d) of SN~2018hna. The prominent features during the late nebular phase ($\sim$\,256 d) are labelled in Figure~\ref{fig:specplat}D. Feature marked \lq \lq A\rq \rq ($\sim$\,8360 \AA) is unidentified and warrants further investigation.

\begin{figure*}[!ht]
\centering
\includegraphics[scale=0.28]{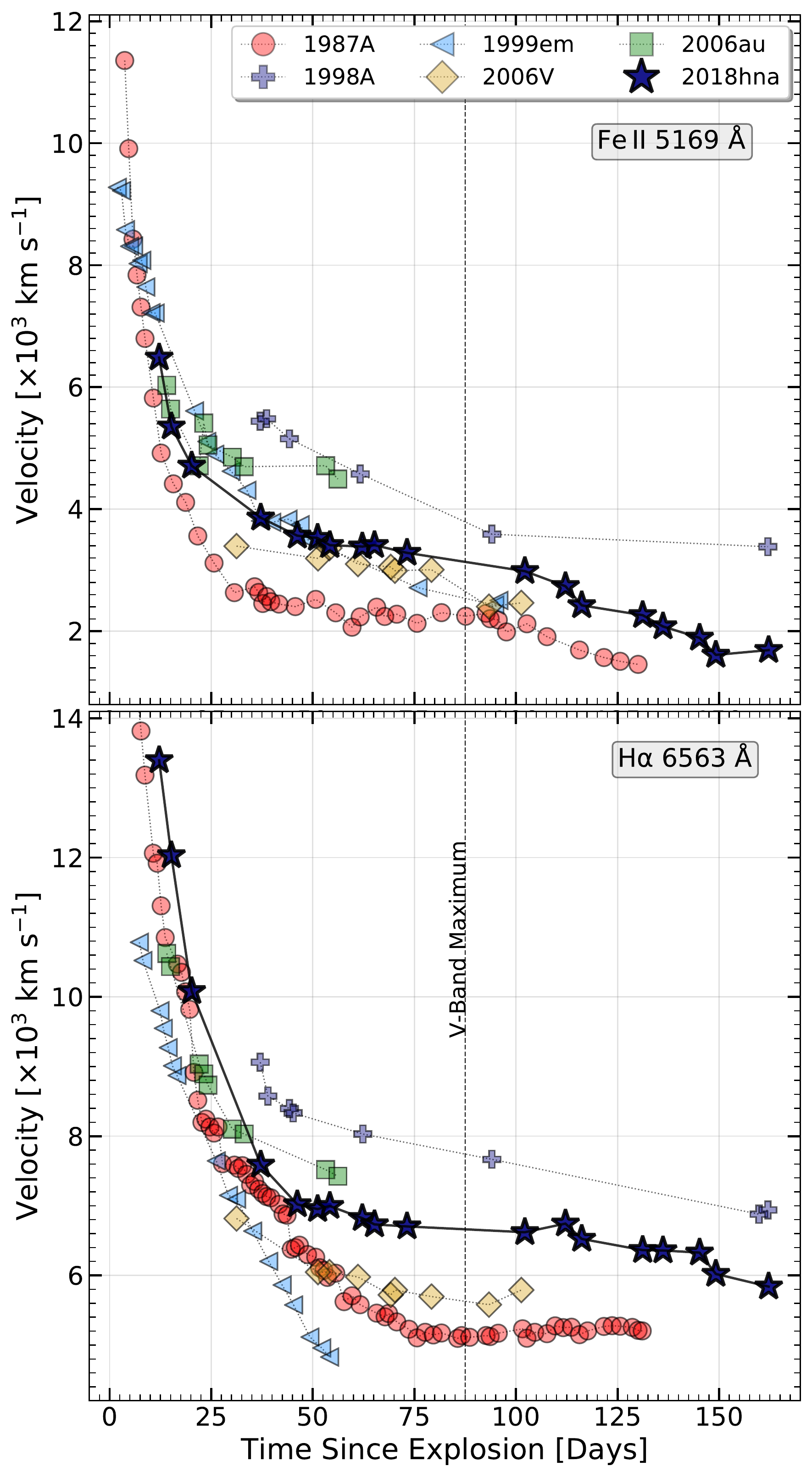}
\includegraphics[scale=0.28]{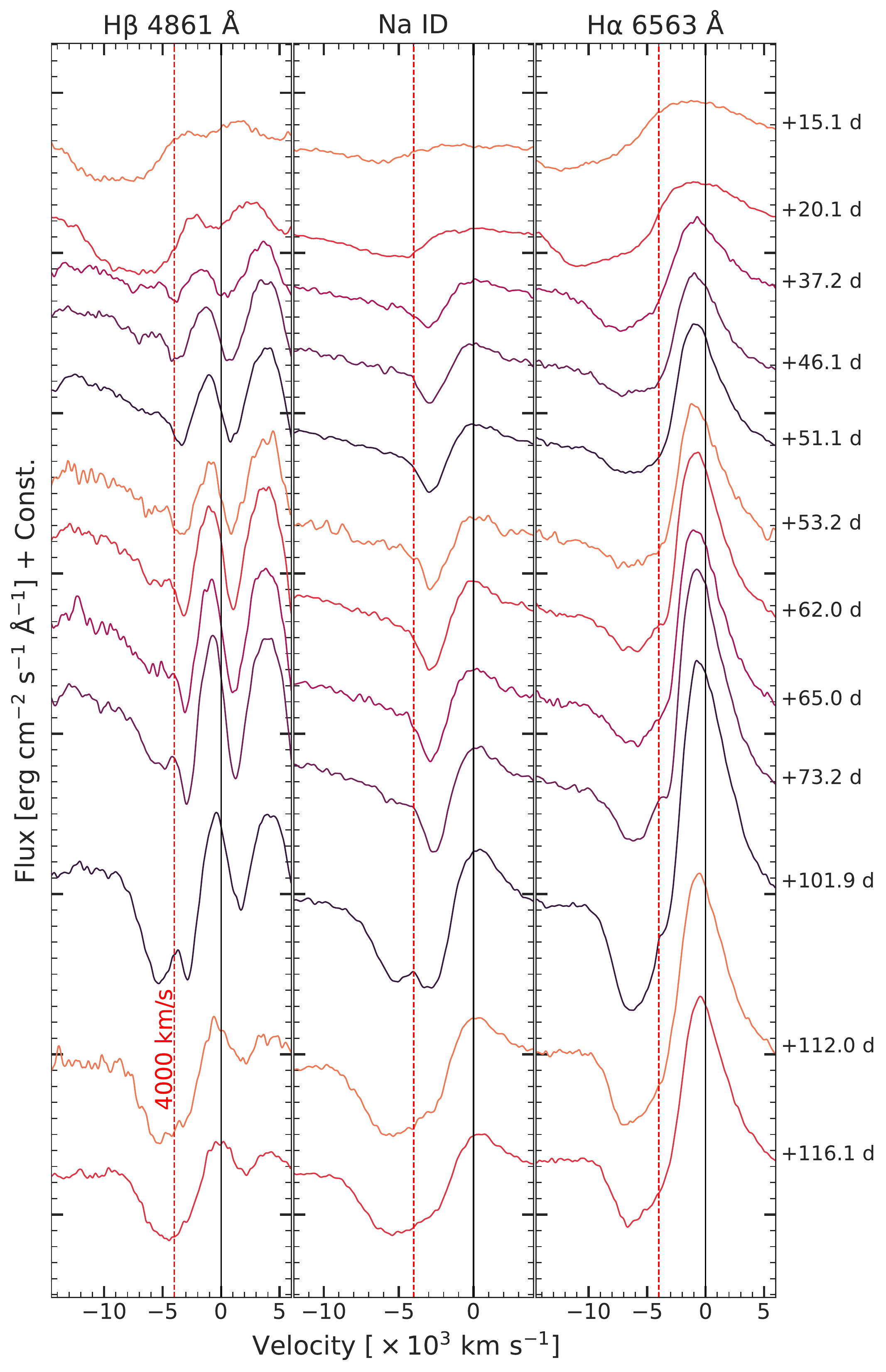}
\includegraphics[scale=0.28]{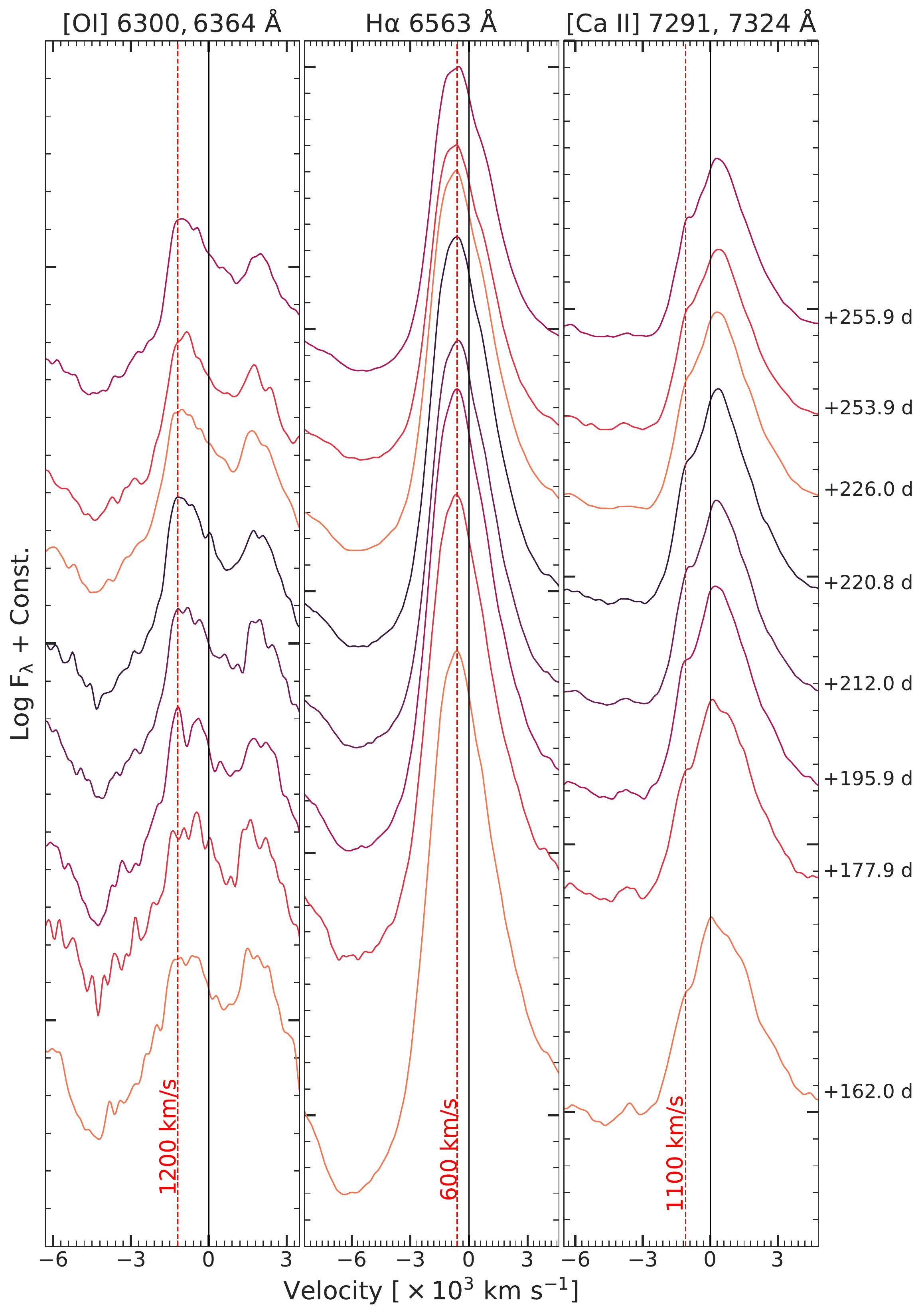}
\caption{\textit{Left panel}: Line velocity evolution of SN~2018hna. \textit{Middle panel}: Evolution of H\,$\rm \beta$, H\,$\rm \alpha$ and Na\,$\rm I$D up until $\sim$\,120 d. \textit{Right panel}: Evolution of \ion{O}{1}, H\,$\rm \alpha$ and \ion{Ca}{2} during the nebular phase.}
\label{fig:photvel}
\end{figure*}

The P-Cygni profile of H\,$\rm \alpha$ in SN~2018hna shows a blue-shift in its emission feature during the photospheric phase, which settles onto a constant value of $\sim$\, 600 $\rm km\ s^{-1}$ during the transition to the radioactive decay phase ($>$\,100 d). The blue-shift has been seen in a majority of Type II SNe (including 1987A-like SNe) and has been explained as arising from the steep density profile of the outer SN ejecta \citep{2014banderson}.

The absorption troughs of H\,$\rm \alpha$, H\,$\rm \beta$ and \ion{Na}{1} D features in the spectra of SN~2018hna show a \lq \lq kink\rq \rq\ (flux excess) beginning $\sim$\,37 d (see Figure~\ref{fig:photvel}). This feature is strongest in the spectrum of $\sim$\,102 d, and disappeared by $\sim$\,112 d. The presence of \ion{Ba}{2} in the troughs of \ion{Na}{1} D and H\,$\rm \alpha$ can possibly explain this feature, although, this line blend should be seen as absorption and not an emission during the photospheric phase. This complex fine-structure was also detected in SN~1987A \citep{1988hanuschik,1989phillips} and comprised of a blue-shifted flux excess and a red-shifted flux deficit, of which only the former is seen in SN~2018hna. The origin of these kinks is likely a result of asymmetry in the line-emitting region. The velocities of these kinks closely follow the photospheric velocity evolution as in SN~1987A and indicate the advent of high-energy radiation from the decay of $\rm ^{56}Ni$ at the photosphere \citep{1989phillips}. 

%-----------------------------------------------------------------------------%
\section{Discussion}\label{sec:discussion}
%-----------------------------------------------------------------------------%

SN~2018hna occurred in the outskirts ($\sim$\,2.5 kpc from the center) of the low-luminosity ($M_B$\,$\sim$\,--17.1 mag) dwarf irregular galaxy (IBm) UGC~07534. The luminosity-metallicity relation of \citet{2004tremonti} yields an oxygen abundance of 8.14\,$\pm$\,0.02 for UGC~07534. This indicates a sub-solar metallicity ($\sim$\,0.3 $\rm Z_{\odot}$) of the host environment of SN~2018hna and is consistent with the occurrence of 1987A-like SNe in late-type galaxies (Sc or later; \citealp{2012pastorello}) having sub-solar metallicity.

\subsection{Cooling envelope emission and explosion parameters}
\label{sec:coolenv}

The early LCs ($<$\,14 d) of SN~2018hna distinctly show adiabatic cooling of the shock-heated SN ejecta. The early emission in the $B$ and $V$ bands shows a rise towards an initial peak. This is a result of the \lq \lq temperature effect\rq \rq\ resulting from the migration of the SN SED into the optical wavelengths while the net bolometric luminosity is still declining \citep{2019fremling}. 
\begin{figure}[!hb]
\centering
\resizebox{\hsize}{!}{\includegraphics{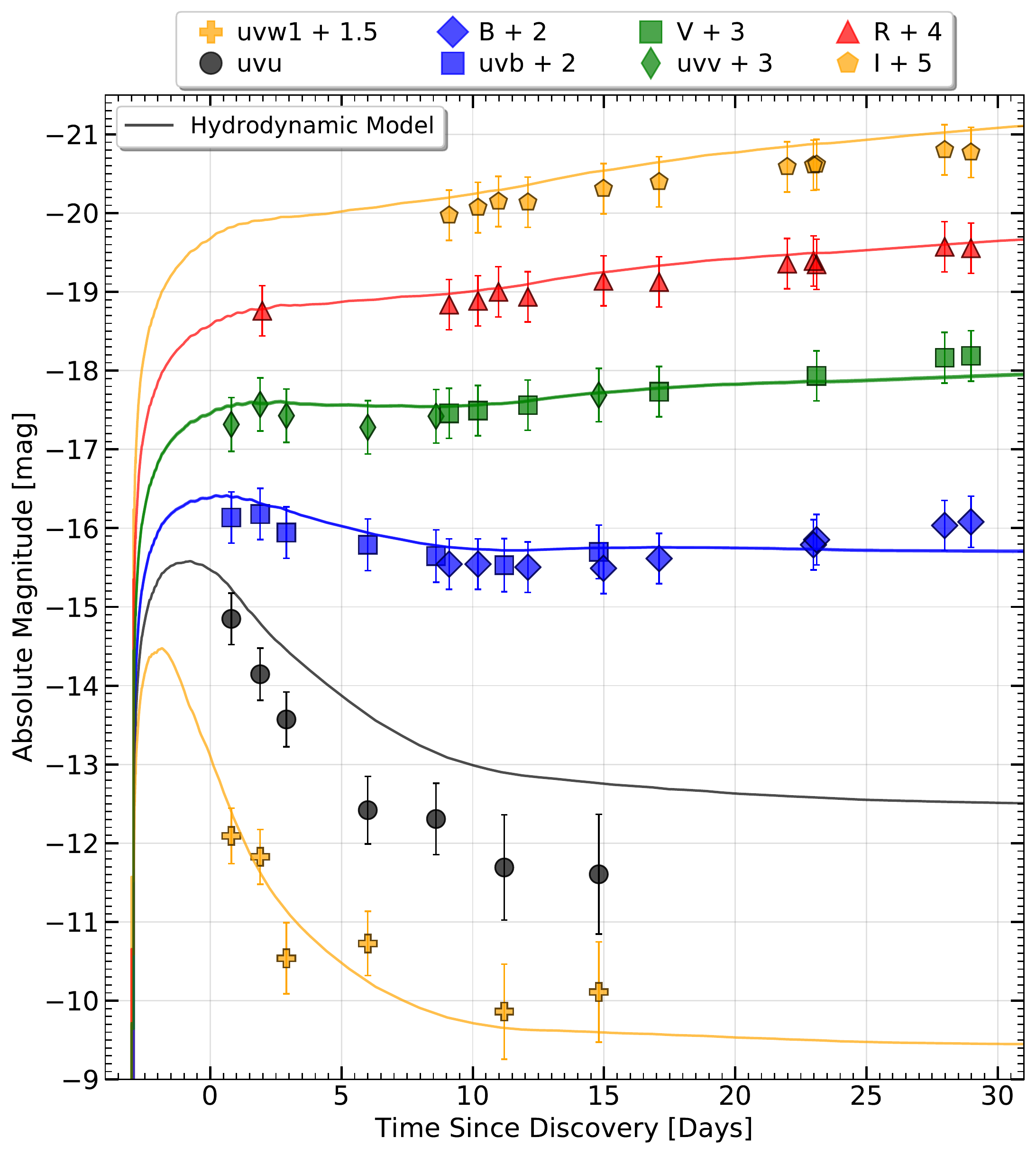}}
\caption{Comparison of early phase \textit{Swift} LCs with the hydrodynamic model.}
\label{fig:modearly}
\end{figure}

Prominent signature of cooling emission is generally seen in Type II (distinctly in IIb) SNe, arising from an extended envelope surrounding their compact core. A similar emission can be seen for a compact BSG progenitor, e.g. 1987A-like events. Since the cooling emission is generally stronger for a more extended and/or more massive envelope, one could use this emission to infer the properties of the H-rich envelope \citep{2010nakar,2012bersten,2014nakar}.

The multi-color light curves of an explosion of a BSG star were computed with the multigroup radiation hydrodynamics code STELLA \citep{2000blinnikov}. The resultant synthetic SED was convolved with the filter transmission function of the respective bandpasses. The multi-color light curves of SN~2018hna during the cooling phase were well reproduced for a star of pre-SN mass $\sim$\,16 $\rm M_{\odot}$ (i.e. an ejecta mass $\sim$\,14 $\rm M_{\odot}$), a radius of $\sim$\,50 $\rm R_{\odot}$ and an explosion energy of $\sim$\,1.7 $\rm \times$\,$\rm 10^{51}$ erg (see Figure~\ref{fig:modearly}). The explosion epoch was constrained to 3 days prior to the discovery i.e JD~2458411.3.

In the case of a compact progenitor, the drop in temperature post the shock breakout reduces once the temperature reaches in the bracket of 6000-8000 K due to the process of recombination. Hence, the luminosity in the redder bands ($VRI$) keeps rising as the temperature falls to this bracket, whereas the luminosity in the UV-bands keeps falling as the wavelength evolves to the Wein's part of the spectrum \citep{2014nakar}. The inset in Figure~\ref{fig:colorbol}A shows the comparison of SN~2018hna with SN~1987A during the cooling envelope phase. The luminosity and time-scale of the cooling emission overlaps between the two SNe. This is indicative of similarity in progenitor properties (like radius and the ratio of ejecta energy to mass) between SN~2018hna and SN~1987A. We rule out the delay of the shock breakout due to an immediate CSM, as no observable feature of such interaction is discernible in the spectral sequence of SN~2018hna.

The long rise-time of 1987A-like events can be explained due to the slow diffusion of radiation from the decay of radioactive $\rm^{56}Ni$ through the massive envelope of the progenitor. Hence, the time taken for the radiation to diffuse through the ejecta is used to estimate the mass using the relation from \citet{1979arnett} for radioactively powered SNe. Using $\rm E_{87A}$ = 1.1 $\times$ $\rm 10^{51}\ erg\ s^{-1}$ and $\rm M_{87A}$ = 14 $\rm M_{\odot}$ \citep{2000blinnikov}, diffusion time, $\rm t_d$\,$\sim$\,1.02\,$\rm t_d^{1987A}$ and a similar mean opacity, an ejecta mass of $\sim$\,19.8 $\rm M_{\odot}$ and an $\rm E_{expl}$ of $\sim$\,2.9 $\times$ $\rm 10^{51} erg$ is inferred for SN~2018hna.

\subsection{Comparison with 1987A-like events}\label{sec:comp}

\begin{figure}
\centering
\resizebox{\hsize}{!}{\includegraphics{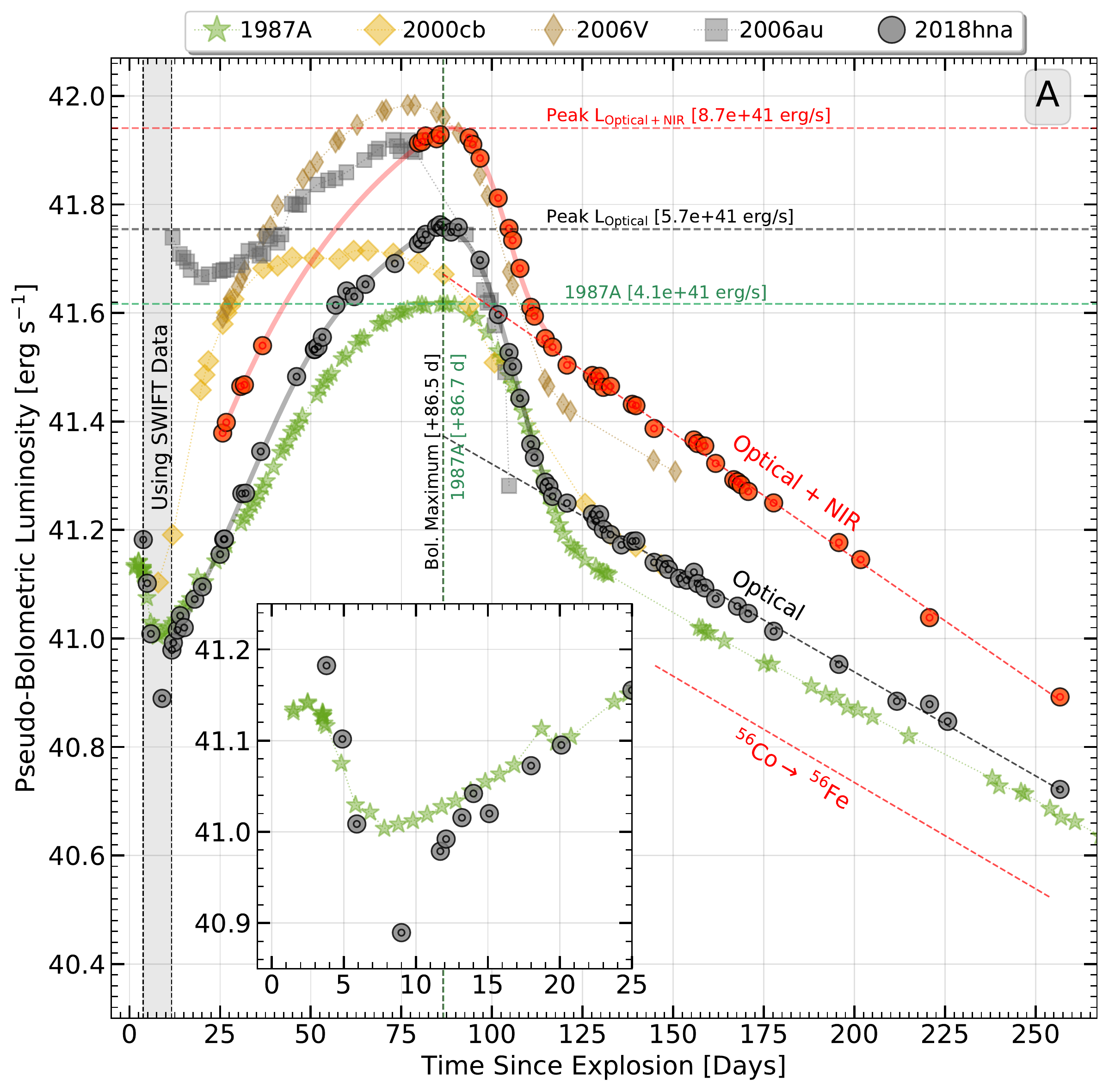}}
\resizebox{\hsize}{!}{\includegraphics{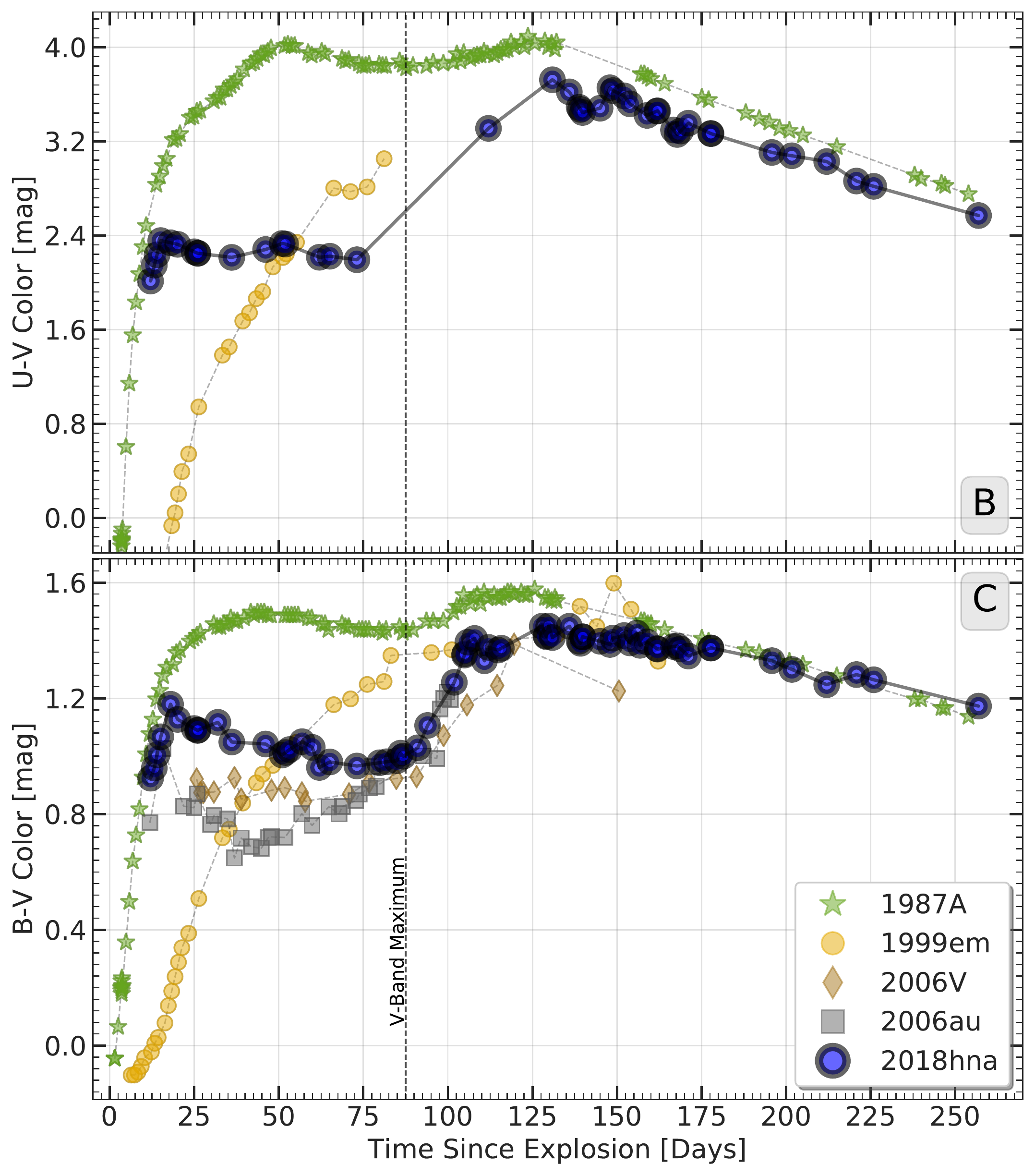}}
\caption{$Panel\ A$: Pseudo-bolometric light curve of SN~2018hna in comparison with 1987A-like events. $Panels\ B\ \&\ C$: \textit{(U--V)} and \textit{(B--V)} color evolution of SN~2018hna in comparison with 1987A-like events and SN~1999em.}
\label{fig:colorbol}
\end{figure}

The \ion{Ba}{2} 6142 \AA\ feature in SN~2018hna was identified as early as $\sim$\,15 d, similar to SN~1987A, but much earlier in comparison to normal Type II SNe (seen $\sim$\,50 d). The strength of \ion{Ba}{2} features in SN~1987A primarily arises from the over-abundance of Barium in the entirety of its hydrogen envelope due to the faster cooling of the SN ejecta resulting from its compact progenitor \citep{1995mazzali}. Also, slow-expanding ejecta are denser and produce stronger \ion{Ba}{2} features in 1987A-like and low-luminosity Type II SNe (e.g. SN~2005cs, \citealp{2009pastorello}). SN~2018hna does not show as strong a \ion{Ba}{2} feature as in SN~1987A due to its relatively bluer colors along with a faster photospheric velocity (50\% faster at \textit{V}-band maximum). This is evident in the spectral comparison in Figures~\ref{fig:specplat}B, C \& D.

The velocity evolution of H\,$\rm \alpha$ and \ion{Fe}{2} $\rm \lambda$5169 feature of SN~2018hna indicates an evolution similar to SN~1987A during the early phase, that subsequently flattens out at a higher velocity (see Figure~\ref{fig:photvel}). This is possibly a result of the higher explosion energy in SN~2018hna, which leads to broader line profiles throughout its evolution. 

The cooling envelope phase in 1987A-like events shows a steep rise in colors as seen in the evolution of \textit{U--V} and \textit{B--V} colors (Figures~\ref{fig:colorbol}B and C). Post the cooling phase, the evolution is almost flat. SN~2018hna, SN~2006au and SN~2006V are bluer compared to SN~1987A. This agrees with the fact that 1987A-like events with bluer colors tend to be brighter at maximum \citep{2016taddia}. SN~2018hna is bluer in \textit{U--V} by $\sim$\,1.6 mag and in \textit{B--V} by $\sim$\,0.4 mag. Differences in color can partially be attributed to higher line-blanketing in SN~1987A (see Figure~\ref{fig:specplat}C) and/or higher $\rm ^{56}Ni$-mixing in comparison to SN~2018hna and other events. Post maximum, the colors turn redder, and are similar to SN~1987A during the transition to the radioactive decay phase (beyond $\sim$\,125 d).

It is seen that the progenitors of 1987A-like events produce higher amount of $\rm^{56}Ni$ in comparison to normal Type II SNe as majority of them have higher ejecta masses \citep{2016taddia}. The degree of $\rm^{56}Ni$-mixing and the net amount of $\rm^{56}Ni$ synthesized helps change the rise to the maximum. A larger $\rm^{56}Ni$ mass increases the time taken to rise to the bolometric peak (rise time) and brightens the peak, whereas a higher degree of $\rm^{56}Ni$-mixing decreases the rise time. 

The optical ($UBVRI$) bolometric luminosity of SN~2018hna is higher than SN~1987A during the maximum and the radioactive decay phase (Figure~\ref{fig:colorbol}A). The bolometric luminosity including NIR contribution ($UBVRIJHKs$) of SN~2018hna is 16\,$\pm$\,5\% brighter than SN~1987A during the radioactive decay phase ($\sim$\,150 d). Hence, the $\rm ^{56}Ni$ synthesized in SN~2018hna is 0.087\,$\pm$\,0.004 $\rm M_{\odot}$. The rise to the bolometric maximum is steeper in SN~2018hna in comparison to SN~1987A and signifies a slightly lower-degree of $\rm^{56}Ni$-mixing in its ejecta. The luminosity at maximum of 1987A-like events is roughly $\sim$\,1.5 times the luminosity from the radioactive decay \citep{2019dessart}, however, in the case of SN~2018hna it is more than twice ($\sim$\,2.5). 

Blue-shifted emission lines in the nebular phase have shown to be an indicator of dust formation in the ejecta of SNe \citep{2003aelmhamdi} as the receding component of ejecta is blocked by the newly synthesized dust \citep{2018asarangi}. Blue-shifted emission of [\ion{O}{1}], H$\rm \alpha$ and [\ion{Ca}{2}] is seen during the nebular phase of SN~2018hna (see right panel in Figure~\ref{fig:photvel}). The peak of these emission lines stays at almost a constant velocity of $\sim$\,1200, 600 and 1100 $\rm km\ s^{-1}$, respectively, throughout the nebular phase. These numbers are consistent with the fact that Oxygen and Calcium are found deeper in the ejecta than Hydrogen. The first overtone of CO emission ($\sim$\,2.3 $\mu$m) was detected in SN~2018hna in the near-infrared spectrum of $\sim$\,153 d \citep{2019rho}. A similar feature was also detected in SN~1987A around $\sim$\,136 d. In all, the signatures above are an indication of dust in the ejecta of SN~2018hna.

%-----------------------------------------------------------------------------%
\section{Summary}\label{sec:summary}
%-----------------------------------------------------------------------------%
A detailed analysis of the photometric and spectroscopic evolution of the SN 1987A-like supernova SN~2018hna is presented in this Letter, based on which the following are inferred:
\begin{itemize}
\item signature of shock breakout is seen in the early phase. SN~2018hna is only the second BSG event caught within a few days from shock breakout;
\item the rise to maximum is slow, similar to other 1987A-like events, with a peak $V$-band absolute magnitude of -16.35\,$\pm$\,0.32 mag;
\item an explosion energy of $\sim$\,1.7-2.9$\rm \times$\,$\rm 10^{51}\ erg$, a BSG progenitor with a radius $\sim$\,50 $\rm R_{\odot}$, and mass $\sim$\,14-20 $\rm M_{\odot}$; and
\item a sub-solar metallicity ($\sim$\,0.3 $\rm Z_{\odot}$) for the host galaxy UGC 07534, in coherence with the low-metallicity of other 1987A-like events.
\end{itemize}
%-----------------------------------------------------------------------------%

%-----------------------------------------------------------------------------%
\section{Acknowledgements}
%-----------------------------------------------------------------------------%

We thank the referee for their positive comments on the manuscript. We thank Masaomi Tanaka for his insightful suggestions. We thank the staff of IAO-Hanle, CREST-Hosakote, Higashi-Hiroshima Observatory and Okayama Observatory who made these observations possible. The facilities at IAO and CREST are operated by the Indian Institute of Astrophysics (IIA), Bangalore. The 0.7m GIT is set up by IIA and the Indian Institute of Technology, Bombay at IAO, Hanle with support from the Indo-US Science and Technology Forum (IUSSTF) and the Science and Engineering Research Board (SERB) of the Department of Science and Technology (DST), Government of India Grant No.IUSSTF/PIRE Program/GROWTH/2015-16. GCA and VB acknowledge the SERB-IUSSTF grants for the same.

DKS and GCA acknowledge DST/JSPS grant, DST/INT/JSPS/P/281/2018. HK thanks the LSSTC Data Science Fellowship Program, which is funded by LSSTC, NSF Cybertraining Grant \#1829740, the Brinson Foundation, and the Moore Foundation. PVB's work on finding parameters of SNe is supported by the grant RSF 18-12-00522. SB is supported by the grant RSF 19-12-00229 in his work on developing codes modeling the radiative transfer in SNe. K.M acknowledges support by JSPS KAKENHI Grant (18H05223) for the initiation of the SN follow-up program with the Seimei telescope. This research has been supported in part by the RFBR-JSPS bilateral program.

This research made use of \textsc{RedPipe}\footnote{\url{https://github.com/sPaMFouR/RedPipe}}, an assemblage of data reduction and analysis scripts written by AS. We acknowledge Wiezmann Interactive Supernova data REPository\footnote{\url{https://wiserep.weizmann.ac.il}} (WISeREP; \citealp{2012yaron}) and ESA Gaia, DPAC and the Photometric Science Alerts Team\footnote{\url{http://gsaweb.ast.cam.ac.uk/alerts}}.

$Facilities$: HCT (HFOSC), GIT, KT (HOWPol, HONIR), ST (KOOLS-IFU), Swift (UVOT)

\software{STELLA, Astropy (v3.1.2), SciPy (v1.3.0), Matplotlib (v3.1.0), Pandas (v0.24.2), PyRAF (v2.1.14), Seaborn (v0.9.0)}

%-----------------------------------------------------------------------------%
\bibliographystyle{AASJournal}
\bibliography{_Reference}

\end{document}